\documentclass[referee]{aa} 
 
\newcommand{\grs}{GRS~1758$-$258}
\newcommand{\unoe}{1E~1740.7$-$2942}

\newcommand{\grp}    {${\rlap.}^{\circ}$}
\newcommand{\pri}    {${\rlap.}^{\prime \prime}$}
\newcommand{\rl}     {${\rlap.}^{s}$}

\newcommand{\ltsima} {$\; \buildrel < \over \sim \;$}
\newcommand{\simlt}  {\lower.5ex\hbox{\ltsima}}            
\newcommand{\gtsima} {$\; \buildrel > \over \sim \;$}
\newcommand{\simgt}  {\lower.5ex\hbox{\gtsima}}            

\usepackage{graphicx}
\usepackage{txfonts}
\usepackage{natbib}
\bibpunct{(}{)}{;}{a}{}{,} 

\begin{document} 

  \title{The precessing jets of \object{\unoe}}

   \subtitle{}

   \author{Pedro L. Luque-Escamilla\inst{1,4}
         \and
          Josep Mart\'{\i}\inst{2,4}
          \and
          Jos\'e Mart\'{\i}nez-Aroza\inst{3,4}
          }

   \institute{Departamento de Ingenier\'{\i}a Mec\'anica y Minera, EPSJ, Universidad de Ja\'en, Campus Las Lagunillas s/n, A3-008, 23071 Ja\'en, Spain\\
             \email{peter@ujaen.es}
            \and
             Departamento de F\'{\i}sica, EPSJ, Universidad de Ja\'en, Campus Las Lagunillas s/n, A3-420, 23071 Ja\'en, Spain\\
             \email{jmarti@ujaen.es}
             \and
             Departamento de Matem\'atica Aplicada, Universidad de Granada, Campus de Fuente Nueva, 18071 Granada, Spain\\
             \email{jmaroza@ugr.es}
             \and
             Grupo de Investigaci\'on FQM-322, Universidad de Ja\'en, Campus Las Lagunillas s/n, A3-065, 23071 Ja\'en, Spain\\
             }

   \date{Received Month xx, 2015; accepted Month xx, 2015}

  
  \abstract
   {The source \object{\unoe}\ is believed to be one of the two prototypical microquasars towards the Galactic center region 
   whose X-ray states strongly resemble those of \object{Cygnus X-1}. 
   Yet,  the bipolar radio jets of \object{\unoe}\ are very reminiscent of a radio galaxy.  The true nature of the object
   has thus remained  an open question for nearly a quarter of a century.}
   {Our main goal here is to confirm the Galactic membership of \object{\unoe}\ by searching for morphological changes of its extended radio jets in human timescales.
   This work was triggered as a result of recent positive detection of fast structural changes in the large-scale jets of the very similar source \object{\grs}.}
   {We carried out an in-depth exploration of the Very Large Array public archives and fully recalibrated all \object{\unoe}\ extended data sets in the C configuration of the array.
   We obtained and analyzed matching beam radio maps for five epochs, covering years 1992, 1993, 1994, 1997 and 2000,  with an angular resolution of a few arcseconds. }
   {We clearly detected structural changes in the arc-minute jets of \object{\unoe}\ on timescales of roughly a year, which set a firm distance upper limit of 12 kpc. Moreover, 
   a simple precessing twin-jet model was simultaneously fitted to the five observing epochs available. The observed changes in the jet flow 
   are strongly suggestive of a precession period of $\sim 1.3$ years.}
   {The fitting of the precession model to the data yields a distance of $\sim 5$ kpc. This value, and the observed changes, rule out any remaining doubts about the \object{\unoe}\ Galactic nature. 
   To our knowledge, this microquasar is the second  whose jet precession ephemeris become available after \object{SS433}.  
   This kind of information is relevant to the physics of compact objects, since the genesis of the precession phenomenon
   occurs very close to the interplay region between the accretion disk and the compact object in the system.}


   \maketitle
%

\section{Introduction}

The microquasar \object{\unoe}\ was first detected with moderate signal-to-noise ratio  during a Galactic plane survey conducted by the {\it Einstein X-ray Observatory} in soft X-rays \citep{1984ApJ...278..137H}.
This object was  later found to be the brightest hard-X source in the Galactic center region
when observed by the coded aperture telescopes ART-P and SIGMA  on board the Soviet GRANAT satellite  \citep{1991A&A...247L..29S}.
The brightness of this object was rivaled only by that of the hard X-ray source \object{\grs}\ discovered by GRANAT a few degrees away from it.
As a result of follow-up monitoring with SIGMA,
\object{\unoe}\ attracted much attention when it was suggested to be a variable emitter of 511 keV $e^- e^+$ annihilation line \citep{1991ApJ...383L..45B}.
This finding was the origin of the "Great Annihilator" nickname for this microquasar.

The source \object{\unoe}\ made further headlines when it was proposed as a highly-absorbed microquasar in the
Galactic center vicinity after the detection of bipolar radio jets emanating from it \citep{1992Natur.358..215M}. 
Only a few months later, \object{\grs}\ was also shown to reveal the same radio jet morphology \citep{1992ApJ...401L..15R}, and thus both sources started to be considered
  sort of microquasar twins in the Galactic center region. 
Nevertheless, the absence of their respective optical or infrared counterparts precluded 
confirming their true Galactic nature via classical spectroscopic techniques. 
The main evidence supporting the microquasar interpretation was the extraordinary resemblance of X-ray spectral properties
with those of the well-known black hole candidate \object{Cygnus X-1} \citep[see, e.g.,][]{1999ApJ...525..901M}.
In this context, the presence of radio jets could also be interpreted as
having been produced in ordinary radio galaxies. Alternatively,
assuming a Galactic center distance of 8.5 kpc, 
the arc-minute angular dimensions of the radio jets translated into a parsec scale linear size of the
collimated jet flows.

In contemporary times, the infrared counterpart of \object{\grs}\ has been confidently identified as a variable, point-like source whose spectral energy distribution
agrees with a microquasar system harboring a low-mass stellar companion \citep{2014ApJ...797L...1L}. 
For \object{\unoe,}\ only a candidate counterpart has been reported in the near-infrared based on astrometric coincidence with the X-ray and radio position
of the microquasar central core \citep{2010ApJ...721L.126M}. Accounting for the high interstellar absorption towards this source, the observed magnitudes agree
 with a high-mass X-ray binary  (HMXB) or  a low-mass X-ray binary (LMXB)  microquasar, depending on the near-infrared emission being dominated or not by stellar emission and, of course, the assumed distance.
Nevertheless, the apparently extended aspect 
of the infrared counterpart candidate still cast some doubts about the true Galactic origin of \object{\unoe} in case the extension was real and not due to
crowding in the stellar field.

Very recently, \object{\grs}\ has been finally shown to be within the boundaries of the Milky Way. This statement is based on causality
arguments applied to the discovery of major
structural  variations in its relativistic jets and lobes \citep{2015A&A...578L..11M}. The observed changes developed on
timescales of about a decade, likely as a result of hydrodynamical instabilites. 
This finding prompted us to search for a similar
behavior in \object{\unoe,}\ which could also rule out the possibility of dealing with an extragalactic source. Although we have not detected  evidence
of major disruptive events in the Great Annihilator jet flow, noticeably morphological changes in its bipolar jets are certainly observed.
An interpretation based on jet precession is tentatively proposed. In any case,
the findings reported here confirm the marginal evidence for time variations in the \object{\unoe}\ extended jets historically reported by \citet{1994AIPC..304..413M},
and positively settle, at last, the issue of its Milky Way membership. 

\begin{figure}
   \centering
   \includegraphics[angle=0,width=9.0cm]{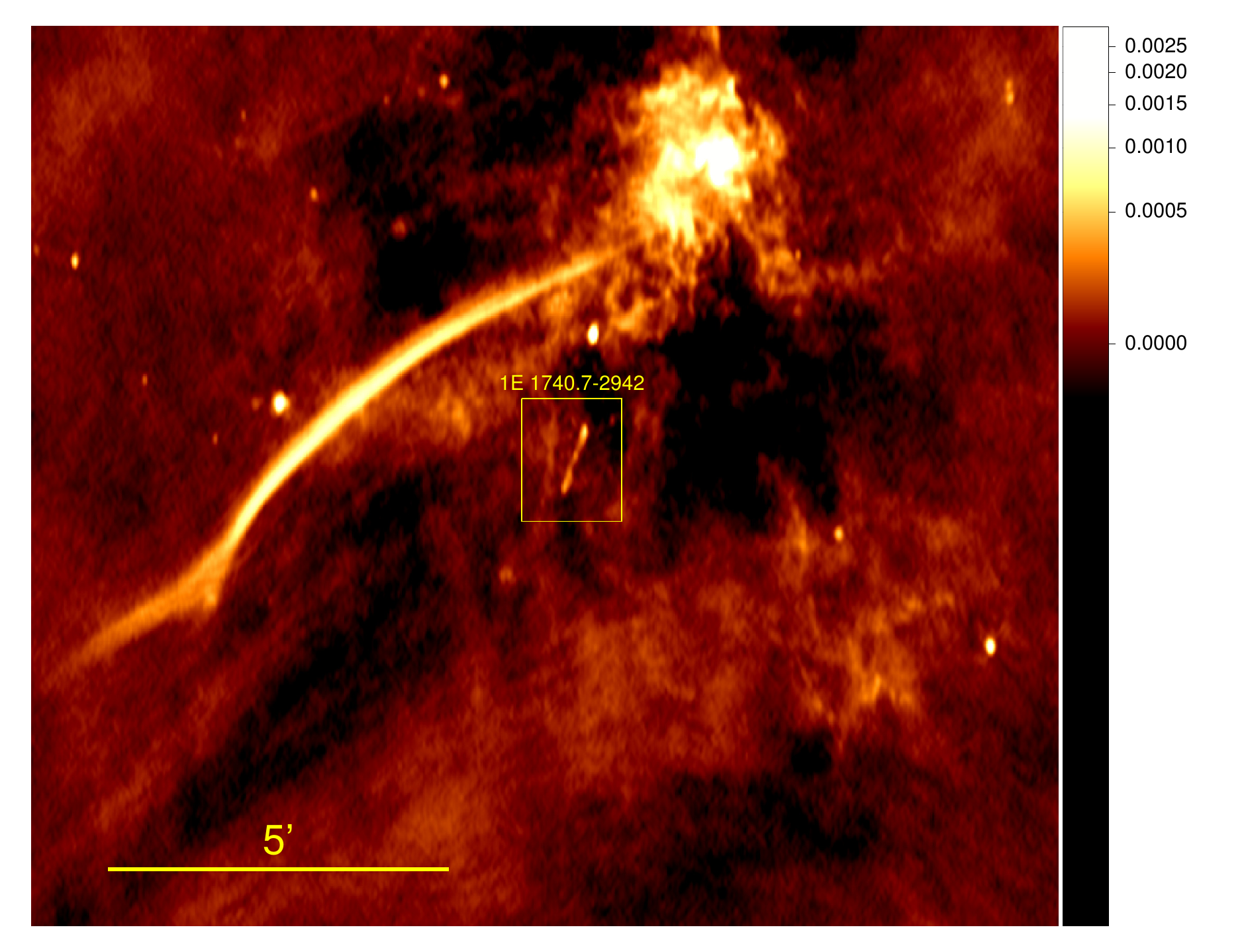}
      \caption{Wide-field, deep VLA radio image of \object{\unoe}\ at the 6 cm wavelength obtained after combining all the visibility data listed in Table \ref{log}.
        The corresponding synthesized beam was  8\pri 16 $\times$  3\pri 47, with  position angle of $-$1\grp 2. 
      The yellow box outlines the region occupied by the microquasar jets that we study with further detail in the rest of this work, while
      the long filamentary structure is the famous Galactic center Snake \citep{1984Natur.310..557Y}.  The horizontal yellow bar gives
      the angular scale. 
      The total integration time amounts to 30 h. The resulting rms noise in this highly confused area of the Galactic center is estimated as 20 $\mu$Jy beam$^{-1}$. 
       Brightness level in a logarithmic scale is given by the color vertical bar in Jy beam$^{-1}$.
       North is up and east is left.
      }
         \label{deep_map}
   \end{figure}

In Fig.  \ref{deep_map} we provide a general view of the \object{\unoe}\ field for illustrative purposes. This figure shows the microquasar
location in a complex region rich in extended radio emission from the Galactic plane. 
The central core of the system lies at the J2000.0 position
$\alpha = 17^h 43^m$54\rl 83 and $\delta =-29^{\circ} 44^{\prime}$42\pri 9 \citep{2010ApJ...721L.126M}.
Other large-scale filamentary features, such as G359.10$-$00.20 also
known as the Galactic center Snake  \citep{1984Natur.310..557Y} are present as well. \object{\unoe}\ is actually located less than one
degree away from Sagittarius A, at an angular distance equivalent to about five 6 cm primary beams of the 
Very Large Array (VLA) antennae.

\section{Observations and data analysis}

We carried out data mining at the public archives of the  
VLA hosted by the National Radio Astronomy Observatory (NRAO) in the US.
The \object{\unoe}\ radio jets are well sampled with an  angular resolution of a few arcseconds with the
$\lambda =6$ cm wavelength in the C configuration of the array.
We thus searched for suitable VLA data sets with this instrumental setup and a reasonable integration time to produce a series of high-fidelity and sensitive maps
with the same angular resolution. Having a good match of the map point spread function, or synthesized beam, is a mandatory condition for 
a meaningful comparison of the radio jet structural changes across different epochs. 
In Table \ref{log}, we list the VLA projects that we finally selected for our study.

Absolute flux density was based on the VLA primary amplitude calibrators \object{1331+305} and \object{0137+331}, while the phase calibrator observed was always \object{1751$-$253}.
Most of the original observations were obtained using the old B1950.0 reference system. They were transformed into
the current standard J2000.0 reference system with the AIPS task UVFIX.

\begin{figure*}
   \centering
   \includegraphics[angle=0,width=18.5cm]{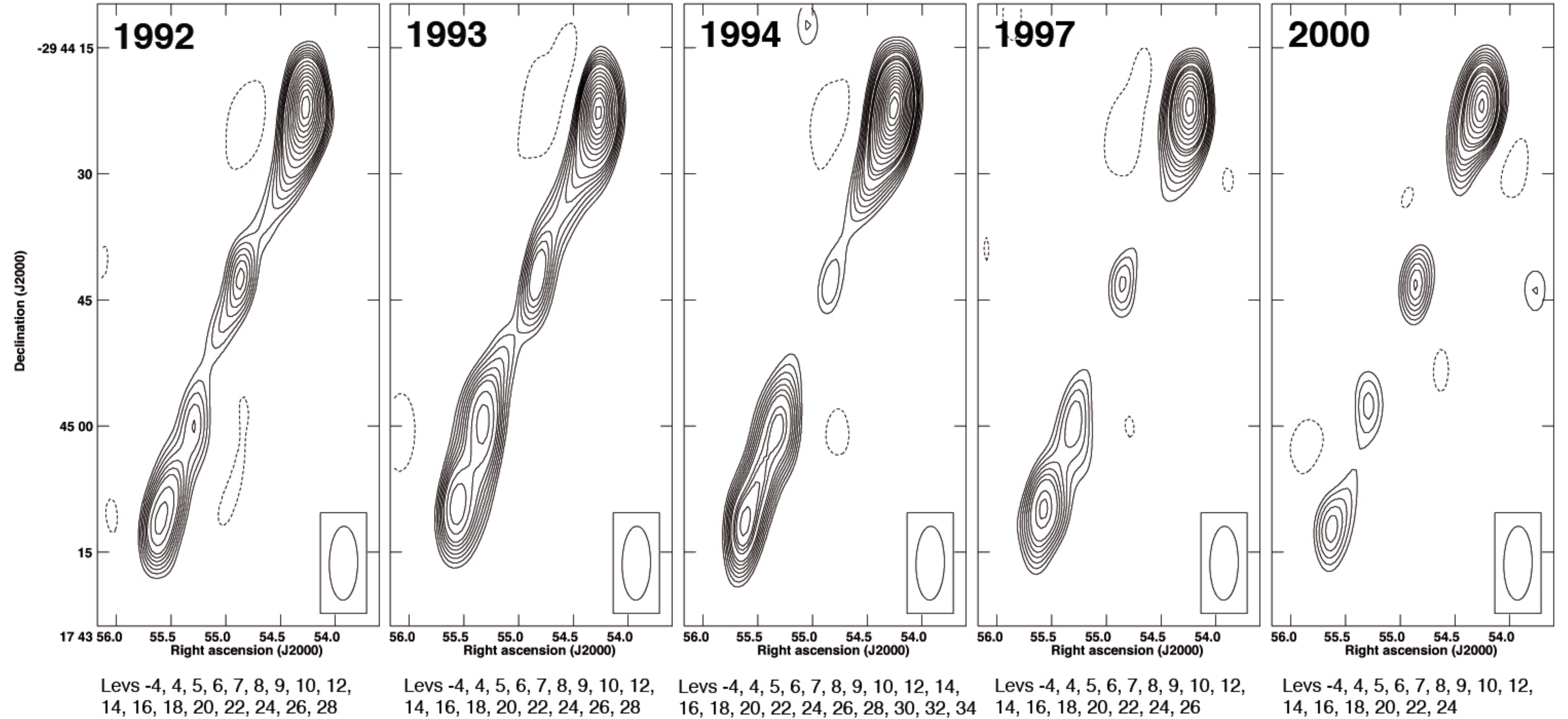}
      \caption{Matching beam contour plots of the \object{\unoe}\ extended radio jets as observed with the VLA interferometer at the 6 cm wavelength (4.8 GHz) over the years 1992 to 1997. The respective rms background noise is 
      20.2, 19.1, 19.1, 21.6, and 21.5 $\mu$Jy beam$^{-1}$. Contour levels in units of rms are given below each frame.  
      The synthesized beam, with a size of 
      8\pri 78 $\times$ 3\pri 38 and $-$0\grp 99 position angle,
      is shown as an ellipse at the bottom right corner on each frame. North is up and east is left. 
      }
         \label{radiomap}
   \end{figure*}

All raw data sets were retrieved and carefully recalibrated using the latest version of the AIPS software package of NRAO. 
Deconvolution of the dirty maps was based on the CLEAN algorithm, as provided by the AIPS task IMAGR. 
 Weighting of the visibilities was pure natural
 (i.e., with a +5 value of the ROBUST parameter in the IMAGR task of AIPS)
 to better enhance sensitivity to the jet arc-minute emission. We also limited the baseline length to larger than
 10 k$\lambda$ to avoid confusion with the Galactic plane emission on much larger angular scales. 
 Figure \ref{radiomap} shows the final result of our multiepoch analysis. 
The same average restoring beam has been used in all observing epochs. 

%
\begin{table}
\caption{Log of VLA  6 cm observations used in this work.}          
\label{log}      
\centering                        
\begin{tabular}{c c c c c}     
\hline\hline                
Project & Array                  &  Observation & On-Source &  Central   \\  
Id.            &  Conf.   &  Date              &  Time (s)  & JD  \\
\hline          
 AM345     &  C   &  1992 Mar 21        &   5280   &     \\       
                 &        &  1992 Apr 09         & 13410   &    2448719  \\
                 &        &  1992 Apr 11         &   5130   &      \\
  AM415    &   C  &  1993  Aug 24      &   8180   &      \\
                 &        &  1993 Aug  27      &   8150   &   2449226   \\
                 &        &   1993 Aug 28      &   8160   &      \\
  AM453    &   C   & 1994  Dec 13       &  11260  &    2449707  \\
                 &         &  1994 Dec 27      &  12690  &      \\
  AM565    &    C  &  1997 Jul   25       &  15070  &   2450657   \\
                 &         &  1997 Jul   28       &  16850  &     \\
  AL511     &    C   &  2000 Apr 07        &    4490  &    2451642 \\   
 \hline
 \hline                                  
\end{tabular}

\end{table}

   \begin{figure}
   \centering
   \includegraphics[angle=0,width=5.5cm]{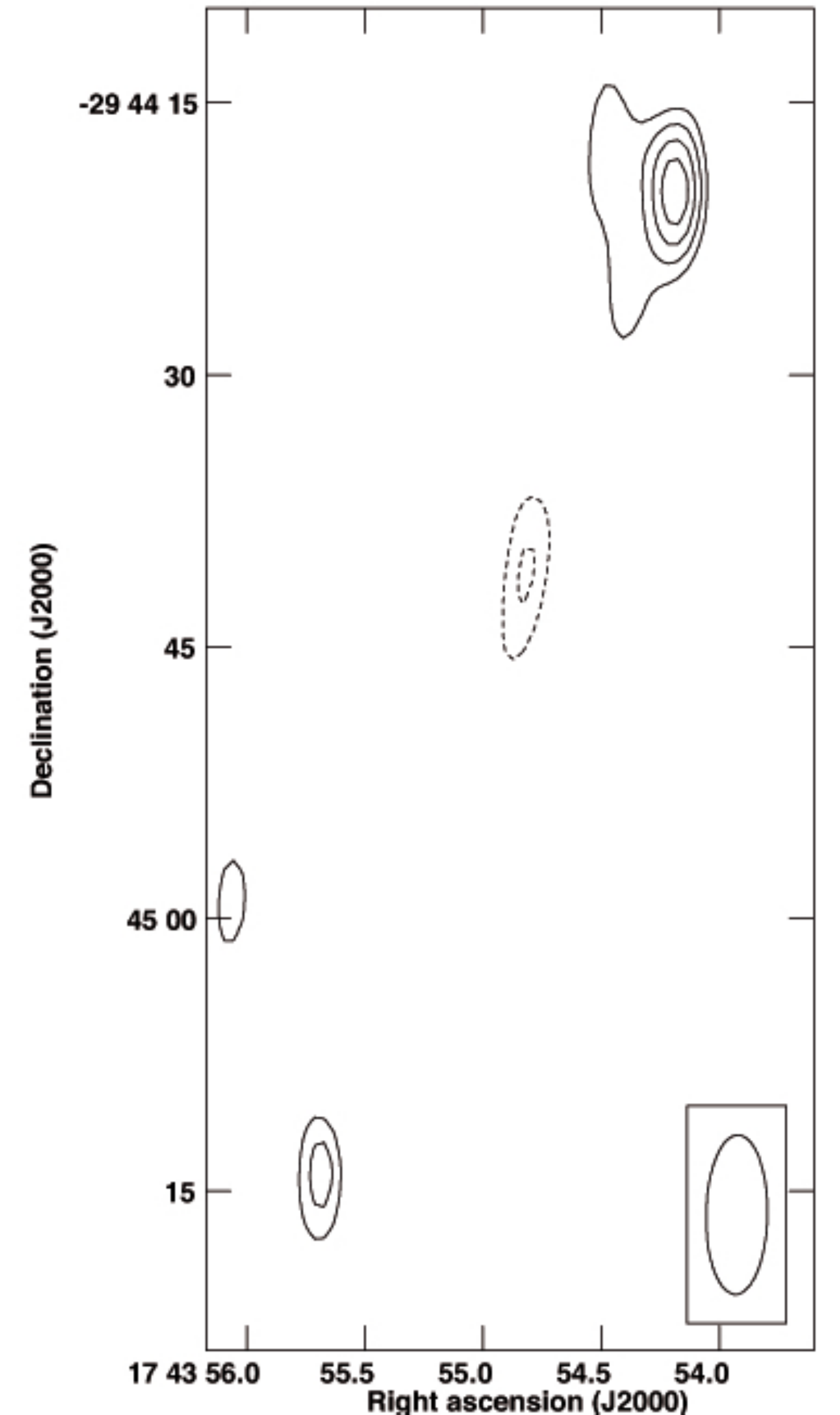}
      \caption{One of the six maps of residuals obtained from the differences of the epochs shown in Fig \ref{radiomap}. 
      This one corresponds to the difference between the 1993 and the 1994 epochs. 
      The rest of residual maps may be seen in the online data. Contour levels are $-5$, $-3$, 0, 2, 4, 6 times the rms noise, which amounts 28 $\mu$Jy. 
      }
         \label{residej}
   \end{figure}

\section{Discussion}

Figure \ref{radiomap} clearly shows that  \object{\unoe}\ jets underwent obvious structural changes in a timescale of years. 
Subtraction of one panel from another (see Fig. \ref{residej} in the main text and Fig. \ref{residuals} in the online material) 
provided strong residuals well above the root-mean-square (rms) noise by a factor up to 6, thus adding confidence to
the reality of the observed variations. 
This is further reinforced as we have independently obtained a residual map for the difference between $1993-1992,$ which is highly compatible with the historical analysis by \citet{1994AIPC..304..413M}. These authors already anticipated the possibility of morphological changes in the \object{\unoe}\ radio lobes, although with less significance than in this work.

\subsection{Distance upper limit}

The most important consequence of the observed changes is an upper limit to the \object{\unoe}\  distance. 
This comes from the timescale ($\tau \simeq 1.3$ yr) of major structural changes
in the bipolar lobes, especially between the 1994 and 1993 observations. The deconvolved angular size
of the strongest northern lobe is about $\theta \simeq 7^{\prime\prime}$. This representative value comes from an elliptical Gaussian fit using the AIPS task JMFIT
on  the Fig. \ref{radiomap} maps, and averaging the major axis over all epochs.
Hence, causality arguments dictate  that the maximum possible distance lies within the boundaries of the Milky Way 
($\sim c \tau / \theta \sim 12$ kpc).
Therefore, we can now be sure that \object{\unoe}\ joins the family of Galactic microquasars just as its twin  \object{\grs}.
  
\subsection{Evidences of precession}

However, the time evolution of the extended radio jets of \object{\unoe}\ does not seem to respond to the same physical arguments based on hydrodynamical instabilities 
invoked in  the \object{\grs}\ case \citep{2015A&A...578L..11M}. 
For \object{\unoe}, the morphology changes  shown in Fig. \ref{radiomap} are instead strongly reminiscent of the precessing radio jets of the well-known microquasar
\object{SS433}  \citep{1981ApJ...246L.141H}. 
It is appropriate to mention here that hints of precession  in \object{\unoe}\ have been previously reported based on X-ray data analysis \citep{2005A&A...433..613D}. 

\subsubsection{A straight line jet?}

Before seriously adopting a precession scenario, one should consider whether
the wavy appearance of the jets could arise from individual point-like blobs 
moving along a straight path, and observed
with the poor resolution of the synthesized beam in the
north-south direction. Including the core, at least four of these blob maxima are well visible in each Fig. \ref{radiomap} contour plot. 
In order to test the quality of a linear fit forced to pass through the central core, we  use a reduced $\chi^2$ approach
as follows.  Let $N_{\rm p}$ be the total number of positions along the jet
measured with respect to the microquasar central core position.
At each  point with  angular coordinates $\left[ \Delta \alpha_i \cos{\delta}, \Delta \delta_i\right]$ ($i=1,...,N_{\rm p}$), 
we assign uncertainties $[\sigma_{\alpha_i}, \sigma_{\delta_i}]$ given by the synthesized beam size divided by the local signal-to-noise ratio of the
jet radio emission. Each point  has two
differential coordinates, thus the total number of observables to be fitted is $2 N_{\rm p}$ (in this case $2N_{\rm p} = 8$).
The reduced $\chi^2$ function is defined as
\begin{equation}
\chi_{\rm red}^2 \equiv 
\frac{
\Sigma_{i=1}^{i=N_{\rm p} } 
\left\{
\left[ 
\frac{\Delta \alpha_i - \Delta \alpha_{\rm fit}}{\sigma_{\alpha_i}}  \right]^2 \cos^2{\delta} 
+
\left[
\frac{\Delta \delta_i - \Delta \delta_{\rm fit}}{\sigma_{\delta_i}}
\right]^2
\right\}
}
{2N_{\rm p} - N_{\rm par}}, 
\end{equation}  \label{chi2}
where $[\Delta \alpha_{\rm fit} \cos{\delta}, \Delta \delta_{\rm fit}]$ are the expected jet differential coordinates at each point from a fitting model
with $N_{\rm par}$ parameters ($N_{\rm par} = 1$ for the slope of the straight line). A fit with $\chi^2$\ltsima 1 is considered very acceptable.

The four condensations quoted above 
 turn out not to be well aligned in all epochs. This is better illustrated in online Fig. \ref{1992}, where the fit is particularly bad, with
 $\chi_{\rm red}^2 = 4.4$. This could suggest a curved jet with some component of blob motion perpendicular to the jet axis, which points again to precession. Moreover, we also tried to check the
assumed compact nature of the blobs by subtracting a point-like source model at each of their positions. 
Full removal of their emission was
not possible unless an extended Gaussian model was used. 
Therefore, the \object{\unoe}\  jet structure seems more likely
to consist of a continuous, extended, and precessing  jet instead of individual, compact blobs all moving along the same straight path.

\subsubsection{Fitting a precessing jet}

To provide additional support to the precession hypothesis, we carried out another  series of fits to the  projected jet paths of \object{\unoe}.
The jet loci were more carefully estimated from the Fig. \ref{radiomap} maps with two different methods. First, by  slicing 
 the jet in the north-south direction pixel by pixel and keeping the
location of maxima in the jet profiles, 
we estimated position uncertainties  as in the previous section and we only kept jet points with signal-to-noise ratio above 5.
Second,  extracting the jet skeleton using the mathematical morphology approach  \citep{opac-b1080653}, 
 allowed us to build two independent
estimates of the jet path, or jet skeletons, whose differences are always less than a small fraction of the synthesized beam. 
To fit these data, we carried out a \object{SS433}-like fit with the same kinematic model developed by \citet{1981ApJ...246L.141H}
several decades ago ($N_{\rm par} = 7$). In contrast to this classical case,  here
 we do not have any additional information from radial velocity changes in the jet flow. The historical 511 keV emission line feature attributed to \object{\unoe}, 
if really originating in the jets, was too broad and too sparsely sampled to assist us in the fit from the Doppler point of view.
Therefore, to our knowledge this is the first time that a multiepoch fit of a relativistic twin-jet kinematic model
is attempted based only on the changing jet path projected onto the celestial sphere.
For comparison purposes, 
 we also fitted a straight line passing through the system's central core 
($N_{\rm par} = 1$ for each epoch).
In both cases, a single $\chi_{\rm red}^2$ value  is computed via Eq. \ref{chi2}, but now with $N_{\rm p} \sim 500$.
It is important to stress here that
all jet paths, on the five different epochs, were simultaneously fitted with the same set of model parameters. 

\begin{figure*}
   \centering
   \includegraphics[angle=0,width=18.5cm]{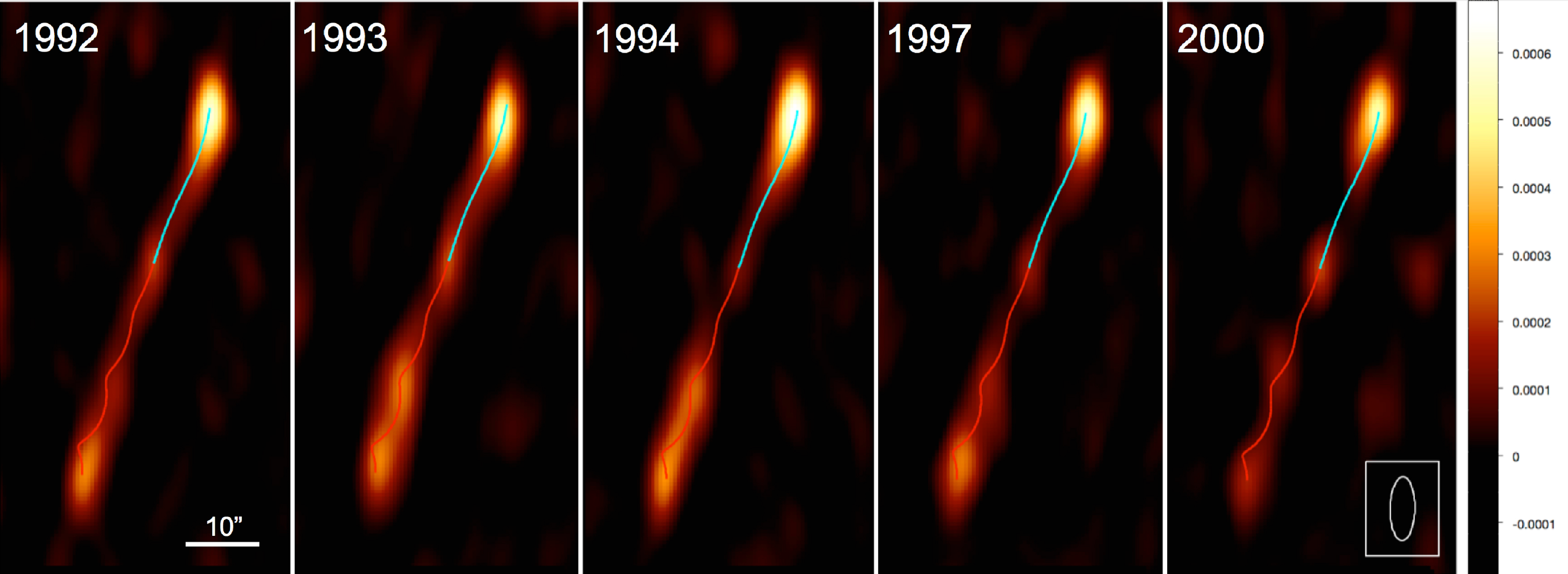}
      \caption{
      Fit of a precessing twin jet model (red receding; blue approaching) using the Table \ref{model} parameters 
      based on the mathematical morphology skeleton that yields $\chi_{\rm red}^2=1.26$.
      The fit is overlayed onto false-color maps of the 1992, 1993, 1994, 1997, and 2000 VLA data sets. 
      The horizontal bar gives the angular scale in arcseconds, and the vertical bar illustrates the brightness scale
      in Jy beam$^{-1}$. The restoring beam used is shown at the bottom right corner. North is up and east is left. 
      }
         \label{time_seq}
   \end{figure*}

The model fits to the jet maxima skeleton are presented in online Fig. \ref{jetfits}.
Again, the simple straight line model does not provide a good description of the data.
Typically, one obtains $\chi_{\rm red}^2$\gtsima 2. The same is true when fitting the 
mathematical morphology skeletons.
On the other hand, significantly better fits are possible when using the precessing jet model. 
After exploring a reasonable range of values in the parameter space, including the distance,
we obtained the best fit  with the jet maxima skeleton with $\chi_{\rm red}^2=0.91$.
This was
followed by the  
$\chi_{\rm red}^2=1.26$ value provided by the mathematical skeleton approach, which is also shown in Fig. \ref{time_seq}.
Their respective seven parameters are  given in Table \ref{model}. A comparison of these parameters does not reveal  differences that are too large.
Varying these parameters by no more than 10\% still provides an acceptable agreement, 
except for the inclination of the jet precession axis ($\pm 5\%$) and the inclination of the approaching jet axis to north ($\pm 1\%$).  

The remarkable good match between the observed jets and the fitted models  suggests that the dominant effect determining 
the extended structure of  the \object{\unoe}\ radio jets is very likely due to precession.

\begin{table}
\caption{Twin-jet kinematic model for \object{\unoe}\ radio jets(*)}          
\label{model}      
\centering                        
\begin{tabular}{lcc}     
\hline\hline                
Parameter &  Mathematical     &   Maxima  \\  
                  &   morphology          &   skeleton   \\
                    &   skeleton            &       \\
\hline      
\hline
Angle of the                & $\psi =$ 1\grp 8           &  $\psi =$ 2\grp 3                  \\
precession cone   &  &  \\
&  & \\
Inclination of the         &  &  \\
jet precession  &  $i = $65\grp 1             &  $i = $63\grp 0                     \\
axis with the l.o.s.                                              &                                     &                 \\
&  & \\
Position angle of the    & $\chi =$ 339\grp 6         &  $\chi =$ 338\grp 5                \\     
jet precession axis  &  &  \\
&  & \\
Precession period                                     &   $P_{\rm p} = 481.8$ d   &  $P_{\rm p} = 485.5$ d             \\
Reference JD(**) & $t_{\rm ref} = 2448303.51$    &   $t_{\rm ref} = 2448307.85$    \\
Jet velocity                                                & $v_{\rm jet} = 0.96 c$          & $v_{\rm jet} = 0.88 c$                 \\
Distance                                                    & $d = 5 $ kpc                         & $d = 5 $ kpc                          \\
 \hline
 Approaching jet (N)                                   & $s_{\rm jet} = +1 $   (fixed)    &  $s_{\rm jet} = +1 $    (fixed)             \\
Receding jet (S)                                        &  $s_{\rm jet} = -1 $    (fixed)  &  $s_{\rm jet} = -1 $      (fixed)               \\
Sense of rotation         &   $s_{\rm rot} = +1$   (fixed)  &  $s_{\rm rot} = +1$      (fixed)            \\
 (counterclockwise)  &  &  \\
 \hline       
 \hline
 Goodness of fit   &    $\chi_{\rm red}^2 = 1.26$  &             $\chi_{\rm red}^2 = 0.91 $ \\
 \hline
 \hline               
\end{tabular}
~\\
(*) Adapted from \citet{1981ApJ...246L.141H}.\\
(**) Corresponding to zero precession phase.\\

\end{table}

\subsubsection{Implications of the precession model}

Our analysis has to be considered as an exploratory but instructive exercise
while waiting for additional epochs to be obtained. Although precession appears to be well established, caution is still
advisable about its true period because the range
of precession phases currently sampled by available data is still very small ($\sim 10\%$ of precession cycle).
The suggested $5$ kpc distance to \object{\unoe}, if correct, comes as a striking result, and not only
because of its consistent agreement with the upper limit derived above. This value would put
 \object{\unoe}\ much closer than usually agreed. 
 Forcing the distance to remain inside the range 8-12 kpc 
 our best fit was obtained for 8 kpc, but it was not very good
  ($\chi_{\rm red}^2=1.23$ using maxima skeleton).
 The rest of fitted parameters did not change significantly, except for a slightly less relativistic jet velocity ($v_{\rm jet} = 0.75c$) and a higher inclination.
  In this case, it is remarkable that the preferred inclination was $>90^{\circ}$, thus implying that the fainter southern jet would be the approaching one.
  The fits also appeared visually too wavy in the northern jet, in contrast to the observed  smoother jet paths in Fig. \ref{radiomap}.
  
  In the absence of further observational data, we  adopt the lowest $\chi^2$ solution as the most likely
   to proceed our discussion.
 At 5 kpc, the arc-minute extension of the jets would then span to a length of 1.5 pc. 
This newly proposed distance is not strictly precluded in any of the analysis  up to now, 
and is in agreement with the arguments already pointed out by \citet{1992A&A...259..205M}. Only some concern could emerge about the possible conflict 
with the extremely high column density
($N_{\rm H}= 1.2 \times 10^{23}$ cm$^{-2}$)  attributed to \object{\unoe}\ from X-ray spectral fits \citep{2002MNRAS.337..869G}. However, one cannot
rule out that an intrinsic source of extinction, such as a slow
and dense stellar wind, is at work.

The newly proposed distance implies consequences for some of the estimations made for the \object{\unoe}\ system in previous works. 
In particular, the absolute magnitude of  the infrared counterpart candidate proposed in \citet{2010ApJ...721L.126M} should be 
revised to $K_s=-3.0$. At 5 kpc, this would be consistent with an early-type B star on the main sequence or a middle-KIII giant star. This last option
is especially interesting because it would render \object{\unoe}\ a system similar to the superluminal microquasar GRS 1915+105, hosting a similar giant
companion star \citep{2001A&A...373L..37G}.
A more common LMXB scenario, with a much fainter nondegenerate companion star, still remains  conceivable if the near-infrared
luminosity arises from the ensemble of an accretion disk plus jet, as in \object{\grs}. A better knowledge of the system spectral energy distribution
is required to discriminate among all these options.

The jet in \object{\unoe}\ seems to be highly collimated, with a small precession cone angle of 
few degrees, and a precession period $P_{\rm p}  \simeq 1.3$ yr. 
\citet{1538-4357-578-2-L129} found two super-orbital periods for \object{\unoe}\,, the main period of $\sim 600$ d, and the other of $\sim 490$ d (or 1.34 yr).
Given their remarkable proximity, it is very tempting to associate this second super-orbital period with the precession cycle proposed here.

\citet{1998MNRAS.299L..32L} derived a theoretical relation between orbital and precession periods in X-ray binary systems. 
Assuming a system with small orbital inclination and mass ratio between star and compact object 
in the range $\mu = 0.2$-$20$, the precession to orbital period ratio is predicted to be in the range $P_{\rm p}/P \sim 18$-$40$. From our precession period, 
and the orbital period $P = 12.7$ d obtained by \citet{1538-4357-578-2-L129}, this ratio becomes about 37 in our case.
This value  would point to a mass ratio close to $\sim 0.2$. 
Let us further consider plausible values of
a $\sim 1 M_\odot$ mass 
for the companion and $\sim 5 M_\odot$ mass for the black hole compact object. 
The semimajor axis of the orbit would then be $\sim 0.36$ AU, while the Roche lobe radius would amount $\sim 25\%$ of this value \citep{1983ApJ...268..368E}. 
Under these kinds of assumptions, the \object{\unoe}\ companion would need to be a giant star to fill such a Roche lobe. 
This would be in agreement with the dereddened absolute magnitude quoted above consistent  with a KIII star. 
Alternatively, in the case of a B-type companion, the mass transfer should most likely
proceed via the stellar wind.

Bulk jet flow velocities close to the speed of light $c$ have been inferred in several X-ray binaries 
such as GRS 1915+105  \citep{1994Natur.371...46M}.  
Thus, the high values of $v_{\rm jet}$  obtained here for \object{\unoe}\ do not come as a  surprise. 
The only relevant aspect is that our fits suggest that the jet flow remains ballistic and highly relativistic until decelerated
very close to the hotspots.
From standard equipartition arguments applied to the VLA maps, the total energy content in the lobes is estimated to be  $\sim 5 \times 10^{43}$ erg
with a magnetic field  $B \sim 10^{-4}$ G.
Assuming steady, continuous, ballistic jets, the obtained velocity
allows us to estimate that the radio lobe particle content is renewed on a timescale of $\sim 10^3$ d,  i.e., the jet travel time.
The corresponding total energy output from the central core is then estimated
to be $\sim 10^{36}$ erg s$^{-1}$. This number is of the same order as in \object{SS433} \citep{1981ApJ...246L.141H} and slightly less than
estimated in the \object{Cygnus X-3} case \citep{2008A&A...479..523S}.

\section{Conclusions}

An in-depth reanalysis of selected data sets of \object{\unoe}\ in the NRAO archive in the time interval 1992-2000 has been presented
that sheds new light on the nature of this bipolar jet source. 
Clear morphological variability has been detected in the arc-minute extended radio jet on yearlong timescales that
confirm the suspicions in this sense advanced two decades ago by \citet{1994AIPC..304..413M}.
Difference maps between observing epochs provide very significative residuals, and this evidence 
changes in the system radio lobes with a high level of confidence (up to 6$\sigma$).
This fact clearly rules out an extragalactic origin for \object{\unoe} and ensures its Galactic microquasar nature.
Moreover, from simple causality arguments a safe 12 kpc upper limit to its distance is derived.

In addition, the \object{\unoe}\ jets show a changing wavy appearance from epoch to epoch that is very suggestive of precession. We tentatively fitted
a precessing twin-jet model that surprisingly yielded very promising results. 
A single set of jet parameters 
simultaneously reproduces the observed jet paths in all five observing epochs. 
\object{The source \unoe\ thus}\ becomes  the second microquasar after \object{SS433}
for which an ephemeris of its precessing relativistic jets is foreseen.

Our preliminary exploration of the parameter
space in the precession model favors a distance of about 5 kpc, i.e., significantly closer than the Galactic center and consistent with the previous upper limit.
In addition, the precession period of 1.33 yr provided by our fit is practically coincident
with one of the super-orbital period reported by \citet{1538-4357-578-2-L129} based on X-ray data analysis.

The closer distance, and also the ratio between the precession and orbital period, 
are found to be reasonably consistent with 
\object{\unoe, which is}\  a LMXB microquasar system 
harboring a giant late-type giant companion filling its Roche lobe. Nevertheless, the currently available data does not strictly 
rule out a HMXB alternative scenario with a main-sequence B-type star.

We did not need to account for any jet deceleration  when fitting the data, thus suggesting that the jet flow
remains highly relativistic until very close to the terminal hotspots.
The corresponding timescales for renewal of the energy content in the \object{\unoe}\ radio lobes imply that the central
engine is able to inject $\sim 10^{36}$ erg s$^{-1}$ into the relativistic jet flow. This value compares satisfactorily
with other Galactic microquasars such as \object{SS433} and \object{Cygnus X-3}.

Future interferometric observations of \object{\unoe}\ over the years will allow a more accurate follow-up of the time evolution
of its bipolar radio jets, and consequently enable a really robust fit to the jet kinematical parameters.

\begin{acknowledgements}
The National Radio Astronomy Observatory is a facility of the National Science Foundation operated under cooperative agreement by Associated Universities, Inc.
Authors acknowledge support by grant AYA2013-47447-C3-3-P from the Spanish Ministerio de Econom\'{\i}a 
y Competitividad (MINECO), and by Consejer\'{\i}a de Econom\'{\i}a, Innovaci\'on, Ciencia y Empleo of Junta de Andaluc\'{\i}a 
under excellence grant FQM-1343 and research group FQM-322, as well as FEDER funds.
\end{acknowledgements}



\bibliographystyle{aa} 
\bibliography{references} 
     

\onlfig{


   \begin{figure*}
   \centering
   \includegraphics[angle=0,width=16.5cm]{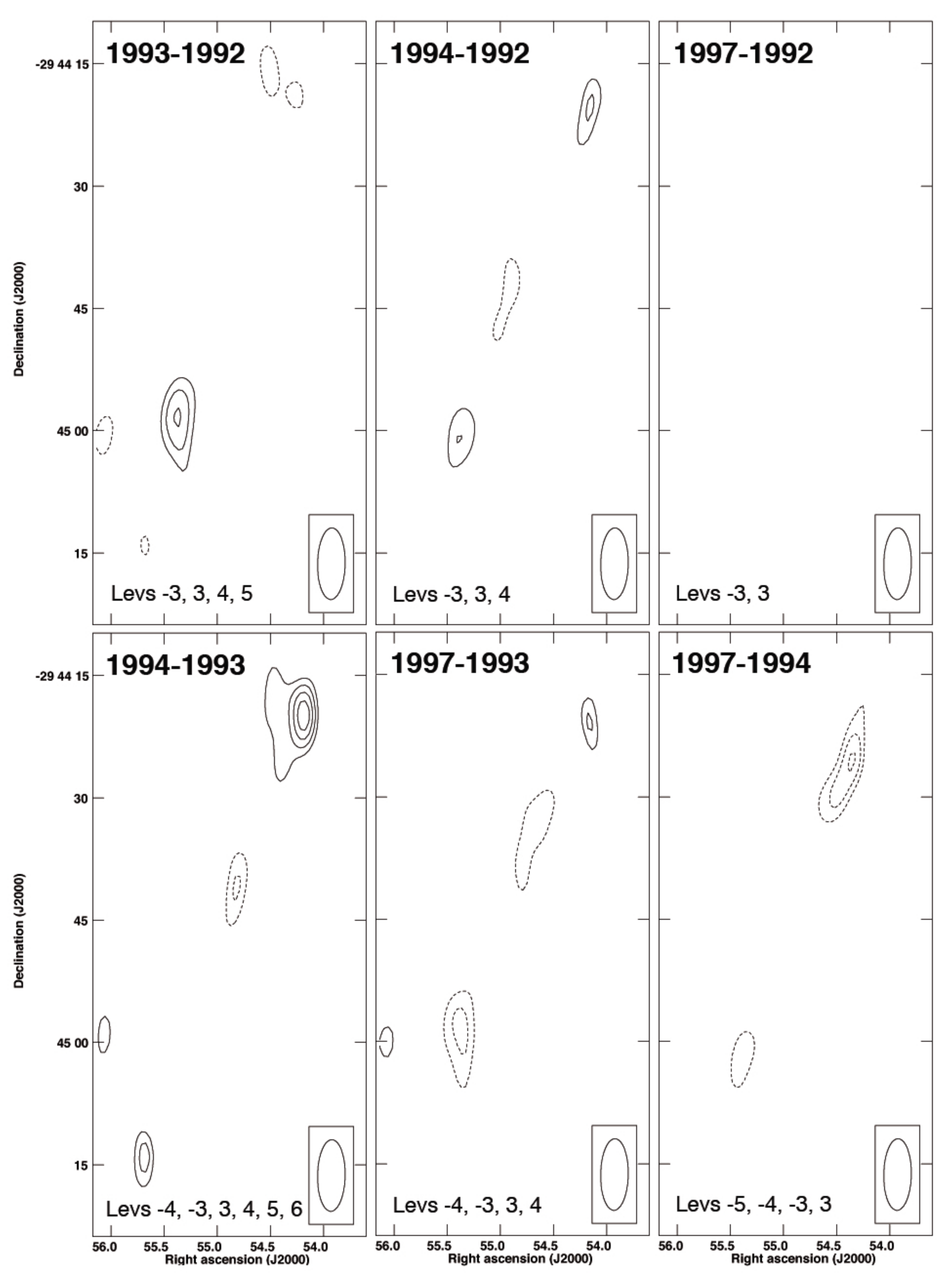}
      \caption{Residuals obtained from the differences between the maps in Fig \ref{radiomap}, as indicated on top of each frame. From left to right and up to down, the rms noise in $\mu$Jy beam$^{-1}$ is 28, 28, 29.6, 28, 29, and 29, respectively. Contour levels are given in units of rms noise  at the bottom of each panel. 
      All residuals tend to be well aligned with the jet position angle.
      The map corresponding to the 2000 epoch has not been used because
      \object{\unoe}\ was not at the phase center and the primary beam corrections would then render the difference maps meaningless.   
      }
         \label{residuals}
   \end{figure*}

     \begin{figure*}
   \centering
   \includegraphics[angle=0,width=12.0cm]{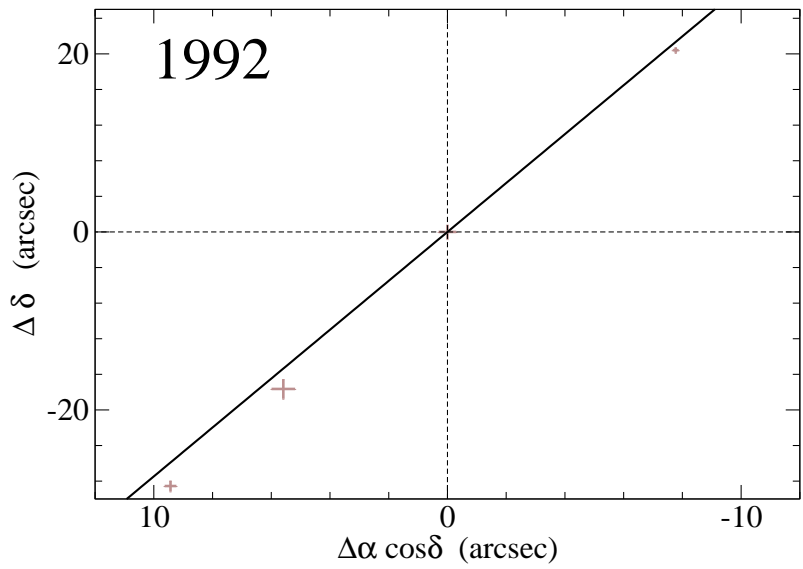}
      \caption{Attempt to fit the 1992 location of the most prominent four condensations in the \object{\unoe}\ jet
        with a simple straight jet line model
      passing through the central core, which results in a rather poor 4.4 value of $\chi_{\rm red}^2$.  
      }
         \label{1992}
   \end{figure*}

 \begin{figure*}
   \centering
   \vspace{-2.0cm}
   \includegraphics[angle=0,width=14.0cm]{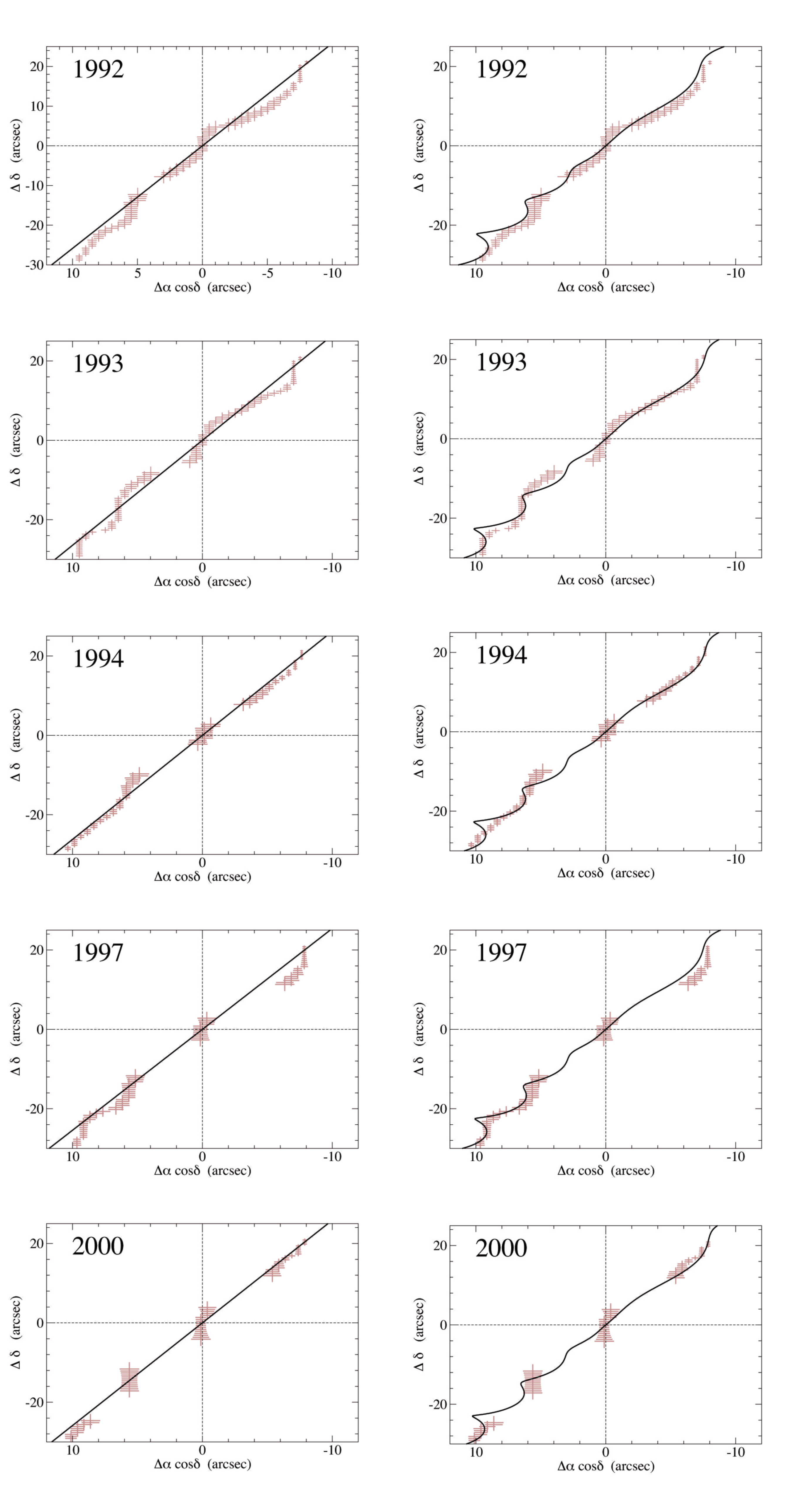}
      \caption{ 
   {\bf Left.}    
      Attempts to fit the 1992 to 2000 observing epochs of the  \object{\unoe}\ radio jets 
       using rectilinear trajectories (thick lines). All of these emanate from the central core 
         located at the intersection of the dashed lines.
      The brown crosses denote the observed jet path based on the maxima skeleton discussed in the text. The horizontal scale has been expanded to better
      show the straight line displacements from the jet flow. The reduced $\chi^2_{\rm red}$ value of the combined five epoch fit
      amounts to 2.2.
   {\bf Right.}   Fits to the 1992 to 2000 observing epochs of the  \object{\unoe}\ radio jets 
based on the \citet{1981ApJ...246L.141H} precessing twin jet model  (thick lines). 
To facilitate an easy comparison, 
       the outline of these plots is the same as in the left panel  except for the fitting curves.
       We used the corresponding model parameter values 
       given in right column of Table \ref{model}. 
       The fit agreement with the observed
       jet paths appears to be significantly better here, with a reduced $\chi^2_{\rm red}$ value as low as 0.91.
      }
         \label{jetfits}
   \end{figure*}

}

\end{document}